\documentclass[12pt]{article}
\usepackage{amsmath}
\usepackage{amsthm}
\usepackage{amsfonts}
\usepackage{amssymb}
\usepackage{eufrak}
\usepackage{graphicx}

\newtheorem{definition}{Definition}
\newtheorem{proposition}{Proposition}

\newcommand{\Lcyc}{[\kern-2pt[}
\newcommand{\Rcyc}{]\kern-2pt]}
\newcommand{\FF}{\mathfrak{F}}

\newtheorem{corollary}{Corollary}[proposition]

\begin{document}

\title{Non-Degenerate One-Time Pad\\ and the integrity of perfectly secret messages}
\author{Alex Shafarenko}
\date{}
\maketitle

\begin{abstract}
We present a new construction of a One Time Pad (OTP) with inherent diffusive properties and a redundancy injection mechanism that benefits from them. The construction is based on interpreting the plaintext and key as members of a permutation group in the Lehmer code representation after conversion to factoradic. The so constructed OTP translates any perturbation of the ciphertext to an unpredictable, metrically large random perturbation of the plaintext. This allows us to provide unconditional integrity assurance without extra key material. The redundancy is injected using Foata's ``pun'': the reading of the one-line representation as the cyclic one; we call this Pseudo Foata Injection. We obtain algorithms of quadratic complexity that implement both mechanisms. 
\end{abstract}

\section{Introduction}

Shannon's concept of perfect secrecy \cite{Shannon} is an important theoretical yardstick by which any imperfect cryptographic solution can be measured. Let us recall what it implies. The scenario is one of Alice sending to Bob a message, which is a bit-string of length $L$, enciphered under a key $K$. Alice requires the probability of a Moriarty deciphering the message to be at most $2^{-L}$ provided that he has no information about the key. Formally, Alice needs the mutual information between the plaintext and the ciphertext to be equal to zero. Then the only way Moriarty can learn the plaintext is by guessing, and the probability to guess a string of $L$ bits correctly in the absence of information about it is $2^{-L}$. The pigeonhole principle quickly leads to the assertion that the length of the key, no matter what cipher is used, cannot be less than $L$ if the secrecy is to be perfect.

One-time pad (OTP), originally proposed by Vernam \cite{OTP}, is the standard method of achieving perfect secrecy: $c_i = K_i\oplus p_i$, $0\le i<L$, where $c_i$ is the ciphertex, $p_i$ the plaintext, $K_i$ the secret key, all three drawn from a set of fixed-length bitstrings with $\oplus$ being modulo-2 addition. The cipher function has to be injective with respect to the plain- and ciphertext, and this is usually strengthened to a bijection to minimise the size of the latter. For example the OTP formula works in both directions due to the fact that  $x\oplus x\oplus y=y$ for any $x,y\in \{0,1\}$. This means that for {\em any} values of $c_i$ there are some corresponding values of $p_i$. An unpleasant consequence of that is that any $c_i$ will decrypt and that Bob is left with no assurance (not even a probabilistic one) that the resulting plaintext has any relationship with what Alice actually sent.

So by itself perfect secrecy is perfectly useless unless messages are guaranteed to arrive untampered with. Without such a guarantee, the cipher only provides security for Alice, but not for Bob. Alice wants her secret not to be available to anyone but Bob, and that is guaranteed. Bob, on the other hand, wants to be sure that the message he has received is exactly the one from Alice, and that is not guaranteed at all.

The standard solution is to add redundancy to the message so that not every ciphertext will decipher to a valid plaintext. This  gives an integrity assurance but requires  additional key material to maintain perfect secrecy for $2^L$ potential messages, since the redundancy makes the message longer than the amount of information in it. 
Normally redundancy and encipherment are decoupled, and the scheme is their composition \[
c = E_K (R_m(p))\,,
\]
where $E: B^{n+m}\times B^{n+m} \to B^{n+m}$ is the encipherment, $R: B^n\to B^{n+m}$ is a redundancy injection, $p\in B^n$ is the plaintext, $c\in B^{n+m}$ is the ciphertext, and $B=\{0,1\}$. The integer $m\ge0$ defines the amount of redundancy injected in the plaintext. If $m=0$, $R$ is the identity function on $B^n$.

For the encipherment to be perfect, there are two requirements. Firstly, the key should be an evenly distributed random bit string of length $n$.  This means that if Moriarty has not obtained a copy of the key, he has no information about it either. Secondly, the
condition \[
(\forall x,y\in B^{n+m})(\exists K\in B^{n+m})E_K(x)=y
\] must hold. Indeed, a given ciphertext should decipher to any given plaintext with an appropriate choice of the key. We call the relation between the key, plaintext and ciphertext defined by $E$ {\em trijective} or a trijection, when it is a bijection between any two of the three values whenever the third one is fixed. 

Since Moriarty has no information about the key, all key values are equally likely and so are all potential plaintexts.  Note that existence does not imply uniqueness, and that generally speaking the key can be longer than the argument of $E_K$, but we will limit ourselves to minimally enciphered messages, i.e. those with $|K|=n+m$, as it is the least length possible for perfect encipherment, see \cite{Shannon}. 

Nor will we consider unprotected redundancy:
\[
c = E_K(p) || D_m(p)
\]
with $E: B^{n}\times B^{n} \to B^{n}$ and $D_m: B^n\to B^{m}$, where $D_m$ is an m-bit digest of the plaintext, since the former leaks information about the latter, requiring the use of a nonce to compensate, again, at the expense of the key length.

The function $R$ is an injection, so the inverse relation is a bijection from $V\subset B^{n+m}$, which is the range of $R$, back to $B^n$, obviously with $|V|=2^n$. The integrity assurance of the scheme comes from the fact that if Moriarty alters the ciphertext $c$ to some $c^\prime$ then the deciphered string $r^\prime=E^{-1}_k(c^\prime)$ will not necessarily belong to $V$. If the nature of the cipher is such that $r^\prime$ is a significant and random deviation from $r=E^{-1}_k(c)=R(p)$ even under the slightest alteration of $c$, then one hopes that the probability of $r^\prime\in V$ (i.e. Moriarty's success probability) is just the cardinality ratio $|V|/|B^{n+m}| = 2^{-m}$, irrespective of the nature of the redundancy injection $R$.

This is the general idea, but the issue is subtle and the details are more than capable of rendering the scheme ineffective. For example, if $E$ is the classical OTP, and $R$ extends the plaintext with $m$ bits of a linear code, then Moriarty is able to produce a $c^\prime\ne c$ that belongs to $V$ with probability 1 at first attempt. All he needs to do is flip a single bit in the data part of the ciphertext and flip the corresponding checksum bits in the enciphered extension according to the code matrix. One could say the difference between $r$ and $r^\prime$ in this case is both small and insufficiently random, even though the idea of using a linear code for redundancy is sound. The problem with the classical OTP is that it is isometric with respect to the Hamming distance: $d_H(c,c^\prime) = d_H(r,r^\prime)$. Worse still, the bit flip operator commutes with the encryption: $k_i\oplus\bar{p}_i = \overline{k_i\oplus p_i}$, making it possible to flip specific bits under the cipher. The transparency of the OTP is a consequence of the fact that the trijection it is based on is {\em degenerate}: its application is position-wise for bit-strings $p$, $c$ and $k$.   

The purpose of the present paper is to establish an integrity control scheme for a perfect cipher. The scheme we seek is one based purely on statistical considerations, without assumptions about the properties of the plaintext, but one which reduces the probability of a successful attack to a value exponentially small in the size of the redundancy. The scheme is not based on computational complexity of any algorithm and so it belongs to the class of {\em unconditional} integrity control schemes, see the related work in Section \ref{sec:relwork}.  

The structure of the sequel is as follows. Section 2 introduces a detailed threat model. Section 3 describes our original non-degenerate one-time pad, where any alteration of a ciphertext component leads to mis-deciphering of not only the corresponding component of the plaintext but also all preceding components thereof, i.e. the cifer exhibits diffusive properties towards the front. In Section 4 we introduce two original bijections: Pseudo Hadamard Transform on Chinese Remainders and First Derivative, which ensure that a ciphertext component alteration influences all plaintext components {\em subsequent} to the one being altered, i.e. they exhibit diffusion towards the back. Section 5 contains our original proposal of Pseudo Foata Injection, which is an effective redundancy mechanism available  in the representation of the plaintext as a permutation of a finite sequence. We also provide some Monte Carlo results showing that the proposed Injection is robust even on its own.

Finally there is are a section on related work and some conclusions.

\section{Threat model}

Perfect ciphers already possess a weak integrity guarantee: clearly a chosen plaintext attack is impossible when the cipher is perfect. We are interested in strong integrity whereby Moriarty is deemed to have succeeded if Bob accepts a message that differs from the genuine message from Alice to any extent at all, down to a single bit. Moriarty is assumed to be able to intercept every copy of the genuine ciphertext message sent by Alice by various routes, modify it, and send it on to Bob. Moriarty has unlimited resources to compute his forgery in short enough time for Bob not to notice the delay. He can also produce an unlimited number of messages that Bob will never refuse to receive, decipher and check the integrity of, until one message succeeds. The scheme is effective if no more than one in $2^{m}$ messages sent by Moriarty does.

So the integrity assurance can not be based on the computational hardness of an algorithm, but solely on the information deficit created by the scheme. 

Note that there is a drastic difference between the confidentiality guarantee based on the security parameter $n$ and the integrity assurance based on $m$. Confidentiality cannot be better than perfect, while $m$ has no theoretical limit. However, in a practical cipher with integrity assurance, it is $m$ that is limited much more drastically than $n$. The reason is that attempts to break imperfect confidentiality are invisible. If the secrecy of a message is based on the computational hardness of the algorithm that breaks the cipher, one has to assume that Moriarty will eventually obtain enough resources to break it. Whereas to get a forgery accepted without breaking the cipher (which is the only option with perfect secrecy), Moriarty has to send many candidate forgeries in the hope to get one of them through. This makes the attack quite explicit, triggering well-understood countermeasures. Also, whereas the computational cost of a forgery is borne by Moriarty alone, and we assume he is not resource limited, message delivery consumes communication resources of both Moriarty and Bob, so Bob is in a position to control the speed of the attack by his input bandwidth.

Accordingly, for a message of, say, 200 bits, the probability of breaking the cipher $2^{-200}$ differs from, say $2^{-64}$ in a practical sense, since one can imagine a quantum computer breaking the latter cipher in reasonable time, but not the former. At the same time if Moriarty needed to send $2^{64}$ 200-bit messages to Bob to  get the latter to accept one forgery, that would be impossible, since Bob would have to receive and process in excess of a billion terabits of data before the probability of Moriarty's success may approach 1. We conclude that the integrity requirements are always moderate (certainly not exceeding the 64-bit level) and that they are independent of the message length. If anything, long messages increase the communication load on Bob, who, according to our threat model has to process every message however received. Consequently, in a real-life situation, the countermeasures will be deployed sooner. 

\section{Nondegenerate one-time pad\label{sec:ndotp}}

Let us switch to a representation where it will be easier to eliminate degeneracy of the cipher and facilitate redundancy injection. This will be done in the following three steps. First we represent the message and the key as members of the symmetric group $S_\nu$ where $\nu!\gtrsim 2^n$. Since no power of two is exactly equal to a factorial the message should be padded with a zero; this does not alter security in a significant way. As far as the key is concerned, it does not have to be binary as it is a random string. 

The second step will be to propose a one-time pad that enciphers the message under the key in a non-degenerate manner. This results in a ciphertext that, if modified slightly (in the sense of some metric), still decrypts to a plaintext that differs from the original a great deal. That difference should be unpredictable and nonlocal, affecting a large proportion of the message. 

\subsection{Lehmer code and factoradic numbers\label{sec:Lehmer}}

A member of $S_\nu$ is a permutation of a length-$\nu$ sequence of symbols, which can be assumed without loss of generality to be numbers taken from the same length range. In the sequel we use the range $[0,\nu-1]$ rather than the more conventional $[1,\nu]$ for our permutations as it simplifies some mappings. To define a specific permutation we can just list the numbers as they appear in the sequence when the base range is permuted. For example for $\nu=4$ this is an identity permutation:
\[
0\ 1\ 2\ 3 
\] 
and this is a cyclic one:
\[
1\ 2\ 3\ 0\,. 
\] 
The above is called {\em one-line notation}. A well-formed permutation in this notation must use all numbers in the range $[0,\nu)$ exactly once, which is a nonlocal constraint, making it difficult to generate random, evenly distributed permutations. The notation constraint can be made completely local by using the Lehmer code\cite{Lehmer-code} as follows.

Let us arrange $\nu$ cells marked $0,1,\ldots,\nu-1$ left to right, and let us place symbols in them in ascending order according to the permutation. Following the above example of a cyclic permutation, we place symbol 0 in position 3, symbol 1 in position 0, etc. In doing so, we will note the distance between the leftmost cell and the cell we are putting the symbol in, while {\em skipping over any non-empty cells}.
So when we place symbol 0, we note distance 3, and when we place symbol 1, we note distance 0. However when we proceed to placing symbol 2 in position 1, the distance will be 0, not 1, since cell 0 is non-empty and we must skip over it. The result is:
\[
3\ 0\ 0\ 0\,. 
\] 
The above representation is the {\em Lehmer code}. Each of the four numbers is completely independent from the rest, and can vary from 0 to $\nu-1-P$, where $P$ is the cell mark. So the first number must be in the interval from 0 to 3, inclusive; the second, 0 to 2, accounting for the fact that one cell is non-empty; the third one, from 0 to 1, since two cells have been used; and finally the last one must be zero: when we have placed all symbols bar one, the last symbol goes in the last empty cell, which has distance 0 from the left as we skip over the rest. Note that for this reason the last element of a Lehmer codeword is {\em always} 0.   

The mapping of permutations onto Lehmer codewords is obviously bijective and the conversion from one-line notation to Lehmer is of no more than quadratic complexity.  Clearly there are as many distinct Lehmer codewords as there are permutations: 
\[
\nu\times(\nu-1)\times\ldots\times1 = \nu!
\] 
To generate evenly distributed random permutations, one only needs to place random numbers numbers (evenly distributed  within the appropriate intervals) in all position of the codeword.

Next, to establish a linear order on permutations, we introduce {\em factoradic numbers}\cite{factoradic} as follows. Consider a positional number system where $j$th digit from the right has a place value of $(j-1)!$ and ranges from 0 to $j-1$. Adding 1 to the maximum digit $j-1$ in position $j$ would make the weighted contribution $(j-1+1)\times (j-1)! = j!$ which is the same as the contribution of digit 1 in the next position to the left, so factoradic numbers work in the same way as numbers in any other positional system. 

Since elements of a Lehmer codeword satisfy the range constrain for factoradic digits, there are exactly as many codewords as there are factoradic numbers of the same length. Since two different factoradic numbers represent different quantities, the representation can be used to enumerate permutations. Moreover, it can be converted to/from binary to make it possible to represent a range of binary messages (or keys).

\paragraph{Conversion to factoradic and back.} To convert a number in the range $[0,\nu!-1]$ to factoradic, one cannot apply the conventional method of repeatedly dividing it by the base while keeping the remainders. The reason for it is that the place values are not powers of the base. However, a less convenient method of repeatedly dividing by the place value, while noting the quotient and replacing the dividend by the remainder for the next round, works just as well. In practice one would precompute the binary strings for divisors for all positions in the factoradic number and use shifts and long subtraction, an algorithm of quadratic complexity. Conversion {\em from} factoradic is similar to the conversion to, the only difference being long addition instead of long subtraction which has the same complexity. 

Here is one complete example for $S_5$. We start with conversion of a binary string 10101 ($21_{10}$) to 5-digit factoradic. The place values are (left to right)  
\[
24\ \ 6\ \ 2\ \ 1\ \ 1.
\]
24 is too large, so the first digit is 0, the second is 3 (remainder 3), the third one is 1 (remainder 1) and the fourth one is also 1 (remainder 0). The Lehmer codeword is thus
\[
0\ \ 3\ \ 1\ \ 1\ \ 0.
\]
Reversing the definition of the Lehmer code, we restore the permutation:
\[
0\ \ 4\ \ 2\ \ 3\ \ 1.
\]

\paragraph{Differential properties.} Since conversion from one-line to Lehmer involves ``skipping over'' used cells as we determine distances, the effect of a perturbation of a Lehmer codeword is nonlocal: when we alter a component, we potentially affect the placement of all subsequent symbols. It is easier to understand such perturbations if we switch to an equivalent definition of the Lehmer code. Now we start with the last symbol $\nu-1$ and we make a sequence of one element out of it and record the last component of the Lehmer code as 0. If the symbol $\nu-2$ occurs in the one-line representation before the symbol $\nu-1$, the Lehmer component indexed $\nu-2$ is 0, otherwise 1, and we assemble the corresponding sequence of the two symbols accordingly. Now for any $0\le \nu-i< \nu-2$ as we progress towards the big end: take the sequence assembled so far of symbols $\nu-i+1,\ldots,\nu-1$ and insert the symbol $\nu-i$ in a place that agrees with symbol precedence in the one-line representation. There will be $i$ distinct insertion points since the  sequence so far is of length $i-1$; insertion before the 0th element gives the Lehmer value 0; and insertion {\em after} the last element, $i$.

It is obvious that when the insertion process is finished, the result is the original one-line sequence and the original Lehmer code. The former is due to the invariant we maintained through the steps, and the latter to the fact that we only look at the symbols subsequent to position $\nu-i$, which means that we skip over all symbols from $0$ to $\nu-i-1$. Here is an example of right-to-left conversion from one-line to Lehmer for the permutation 
\[
0\ \ 4\ \ 2\ \ 3\ \ 1
\]
that we have used before:
\begin{center}
\begin{tabular}{|c|c|}
\hline
Sequence&Code\\
\hline
4&0\\
4\ \ 3& 1\ \ 0\\
4\ \ 2\ \ 3& 1\ \ 1\ \ 0\\
4\ \ 2\ \ 3\ 1& 3\ \ 1\ \ 1\ \ 0\\
0\ \ 4\ \ 2\ \ 3\ \ 1& 0\ \ 3\ \ 1\ \ 1\ \ 0\\
\hline
\end{tabular}
\end{center}

\begin{figure}[t]
\begin{center}
\includegraphics[scale=0.75]{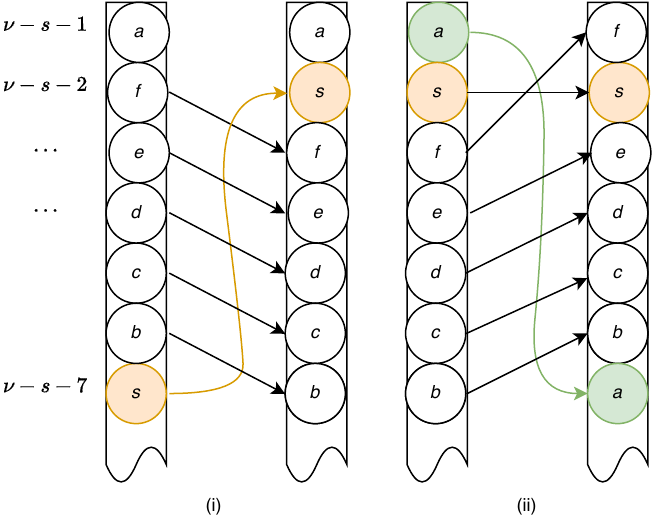}
\caption{Differential properties of the Lehmer code}
\label{fig:lemdif}
\end{center}
\end{figure}

Now we have the necessary tools to discuss the differential properties, namely what happens to the sequence of symbols when a Lehmer component is altered. Consider the illustration shown in Figure \ref{fig:lemdif} case (i). This is a snapshot of the sequence (arranged bottom up) under right-to-left, one-line-to-Lehmer conversion when it reached a symbol $s$. The value of its corresponding Lehmer component is such that $s$ is placed at $\nu-s-7$. We assume that \[
s<a<b,c,d,e,f
\]
with the relationship between the last five symbols not being constrained. Now increase the Lehmer component of $s$ by 5. According to the right-to-left definition of the Lehmer code, $s$ should be deleted at position $\nu-s-7$ and re-inserted at position $\nu-s-2$ in the one-line representation as shown in the figure. The effect of this is equivalent to that of five transpositions: $(s,b)$, $(s,c)$, $(s,d)$, $(s,e)$, $(s,f)$. Of course symbols with values less than $s$ will be inserted anywhere in that diagram, before the conversion is complete. However, those symbols will end up being inserted in the same positions before and after $s$ is moved and the effect of the movement will be the same. The minimum number of transpositions required to reach one permutation from another is a proper metric and is called Cayley distance \cite{Cayley-dist}. It easy to see that offsetting a Lehmer component $w_i$ by some integer $\Delta$, such that $0\le w_i+\Delta <\nu-i$ results in a permutation at a Cayley distance of $|\Delta|$ from the original.

Note that distances induced by subsequent offsets do not accumulate. Looking at diagram (ii) in Figure \ref{fig:lemdif} we find what happens if, subsequent to the first alteration, $a$ moves from position $\nu-s-1$ to position $\nu-s-7$, a jump over 6 places. Symbol $s$ will not move, since it {\em precedes} $a$, and the rest of the symbols are pushed up by one place due to the insertion of $a$. The combined effect of two alterations (of $s$ and of $a$) is that symbols from $b$ to $e$ are not moved, and the Cayley distance of the result from the original sequence is only 2, since the transposition $(f,s)$ followed by $(s,a)$ maps the final permutation onto the initial one. A distance of two is much shorter than either distance to the intermediate stage.

\subsection{Non-Degenerate One-Time Pad (NDOTP)} 

We start with a non-binary degenerate trijection with a free parameter. 

\begin{proposition}\label{prop:elt}
Let $r$ to be a positive number; $0\le p,k,c < r$ an integer plaintext, key and ciphertext, respectively; and $\pi$ an arbitrary cyclic permutation from $S_r$. Then the tripartite relation\footnote{We use square brackets to index a cyclic permutation in one-line notation.} $c = \pi^k[p]$ is a trijection under any choice of $\pi$.  
\end{proposition} 
\begin{proof}
The bijective relation between $c$ and $p$ is due to the fact that any permutation is invertible. 

The cyclic permutation $\pi$ can be visualised as a ring on which all unique nonnegative numbers less than $r$ are placed in some order. The output of the $k$th power of $\pi$ in a given position $j$ along the ring is the number found $k$ steps away from $j$. Clearly, as $k$ goes from $0$ to $r-1$ all members will be encountered and so for any $p$ and $c$ there exists a single $k$ such that $c = \pi^k[p]$.

We conclude that the relation is trijective. 
\end{proof}

\begin{proposition}\label{prop:otp}
Let $\nu\ge2$ and let $0\le p_i,c_i,k_i <\nu-i$ for $1\le i <\nu$ be the Lehmer component of the plaintext, ciphertext and key, respectively, with index $\nu-i-1$. The following recurrence relation in $i$:
\begin{align}
\pi_i &= R_i(\pi_{i-1}, k_{i-1}, p_{i-1}) \nonumber \\
c_i &= \pi_i^{k_i}[p_i] \nonumber
\end{align}
for $i=2,\ldots,\nu-1$ with the initial condition $\pi_1 = 1\ 0$ {\rm (\/}the only cyclic permutation of $S_2${\/\rm)}, defines a non-degenerate trijection between $p$, $c$ and $k$. Here $R_i:K_{i-1}\times \mathbb{N}_{i-1}^2\to K_i$, where $K_i$ is a set of all cyclic permutations from $S_{i+1}$ and $\mathbb{N}_i$ is the set of first $i$ nonnegative integers, is a family of arbitrary surjections\footnote{Surjectivity is not essential, but we require it for statistical properties.} that significantly depend on all their arguments.
\end{proposition}
\begin{proof}
(by induction in $\nu$) The base case $\nu=2$ follows from Proposition \ref{prop:elt} and the fact that a mapping of a single-component Lehmer codeword cannot be degenerate. For the inductive step, assume that the relation for $\nu=\nu_0$ is trijective and non-degenerate. According to Proposition \ref{prop:elt}, the relation between $c_{\nu_0}$, $p_{\nu_0}$ and $k_{\nu_0}$ is also trijective. Consequently the combined, 
$(\nu_0+1)$-component plaintext, ciphertext and key are in a trijective relation. Non-degeneracy follows from the fact that $\pi_{\nu_0}$ significantly depends on all preceding $p_i$. 
\end{proof}

It should be noted that the encipherment introduced by Proposition \ref{prop:otp} is constructive. The decryption process uses $\pi_1^{-k_1}$ and $c_1$ to obtain $p_1$, and for every $i>1$ it calculates $\pi_i$ based on the decrypted $p$'s from the earlier steps. Decryption at step $i$  is achieved by computing
\[
p_i=\pi^{-k_i}_i[c_i]\,.
\]
We call the cipher detailed in Proposition \ref{prop:otp} a Non-Degenerate One-Time Pad (ND-OTP).

The requirement that functions $\{R_i\}$ are surjective comes from the intention to maximise the differential uncertainty. If $c_i$ is perturbed by any small amount amount (down to 1), the corresponding plain text can take any legal value depending on the key. This is in contrast with, say a fixed cyclic permutation, for example $\hat{\pi}_{i}[l]=l+1\mod i$, which would result in 
\[
c_i=p_i+k_i\mod i\,.
\]
Such a choice would be vulnerable to a differential attack whereby Moriarty increments or decrements some Lehmer components of the ciphertext to immediately obtain a plaintext within short Manhattan distance from the original.  

To eliminate such attacks we should also even out the differential uncertainty by demanding that all $\pi_i$ are random and uncorrelated. This is, surprisingly, possible by utilising the true randomness of the key, which is required for perfect secrecy anyway. We can assume that all Lehmer components of the key are evenly distributed within their legal ranges (otherwise the key's entropy would be less than what Shannon's model requires). 

We conclude that the functions $\{R_i\}$ should build a random cyclic permutation $\pi_i$ from a smaller-sized one, $\pi_{i-1}$ 
\begin{enumerate}
\item by a uniformly random expansion, 
\item in such a way that $\pi_i$ depends on the plaintext $p_{i-1}$ in a significant way.
\item in such a way that $\pi_i$ is dissimilar to $\pi_{i-1}$, 
\end{enumerate}

Before we define and illustrate the construction of $\pi_i$, we would like to remind the reader of another standard notation for permutations, {\em cyclic}. In cyclic notation, a permutation is considered to be a collection of independent cycles. Symbols belonging to each cycle are listed between parentheses, in the order that they appear on the cycle, starting with an arbitrary member. For example the following permutation from $S_5$
\[
4\ 2\ 3\ 1\ 0
\]
can be written in cyclic notation as 
\[
(0\ 4)\ (1\ 2\ 3)
\]
A cyclic permutation will have only one bracketed list in cyclic notation and its components can be indexed the same way as in one-line notation, except the origin is not the first symbol (there is no `first' in a cycle), but the symbol `0'. When we need to select a component of a cyclic permutation in cyclic notation, we use double brackets: $\pi\Lcyc 0 \Rcyc = 0$.

Now to points 1--3 above. The first requirement  is easy to achieve by utilising $k_{i-1}$. Permutation $\pi_{i-1}$ is a member of $S_i$, so its symbols run from 0 to $i-1$. Since it is cyclic, there are $i$ distinct positions (after each symbol) in cyclic notation for insertion of the new symbol, $i$, in the cycle. We use $k_{i-1}$, which has the same range, to define that position:
\begin{equation}
\pi_{i}\Lcyc j\Rcyc = \begin{cases} 
\pi_{i-1}\Lcyc j\Rcyc,\; \hbox{if}\; j<k_{i-1}\\
i,\; \hbox{if}\; j=k_{i-1}\\
\pi_{i-1}\Lcyc j-1\Rcyc\; \hbox{otherwise.}
\end{cases}	\label{eq:leftprop}	
\end{equation}
Clearly if all $k_i$ are uncorrelated random numbers uniformly distributed within their legal ranges, then all $\pi_i$ are random and evenly distributed cyclic permutations. We also note that this means that the function $R_i$ is a surjection. However, Equation \ref{eq:leftprop} does not satisfy requirement 2. Consider the following modification:
\begin{equation}
\pi_{i}\Lcyc j\Rcyc = \begin{cases} 
\psi_{i-1}[\pi_{i-1}\Lcyc j\Rcyc],\; \hbox{if}\; j<k_{i-1}\\
i,\; \hbox{if}\; j=k_{i-1}\\
\psi_{i-1}[\pi_{i-1}\Lcyc j-1\Rcyc],\; \hbox{otherwise.}
\end{cases}\,,	\label{eq:lprd} 
\end{equation}
where for all $0\le j<i$
\begin{equation}
\psi_i[j] = \rho_i[j + p_i \mod i]
\label{eq:cut}
\end{equation}
and 
\begin{equation}
\rho_i[j] = \begin{cases}
j/2,\; \hbox{if $j$ is even}\\ 
(j-1)/2+\lceil i/2\rceil,\; \hbox{otherwise.}\\ 
\end{cases}
\label{eq:riffle}
\end{equation}
Here $\rho_i$ is the inverse riffle of a deck of size $i$, and $\psi_i$ is the same after the cut at the distance $p_i$. 

Here are some examples. For $i=5$, this is the cut at $p_i=2$ in one-line notation:
\[
2\ 3\ 4\ 0\ 1.
\]
The inverse riffle of the same length:
\[
0\ 3\ 1\ 4\ 2.
\]
and this is the combined permutation:
\[
2\ 0\ 3\ 1\ 4.
\]

From elementary group theory, Equation \ref{eq:lprd} can be rewritten perhaps more elegantly using the group operation $\circ$ thus:
\begin{equation}
\pi_{i}\Lcyc j\Rcyc = \begin{cases} 
(\psi_{i-1}\circ\pi_{i-1}\circ\psi^{-1}_{i-1})\Lcyc j\Rcyc,\; \hbox{if}\; j<k_{i-1}\\
i,\; \hbox{if}\; j=k_{i-1}\\
(\psi_{i-1}\circ\pi_{i-1}\circ\psi^{-1}_{i-1})\Lcyc j-1\Rcyc],\; \hbox{otherwise}\\
\end{cases}\,,	\label{eq:lprdg} 			  
\end{equation}
to emphasise the fact that the transformation of $\pi$ by $\psi$ is conjugation, but Equation \ref{eq:lprd} suggests a direct implementation in the cyclic representation.

\paragraph{Discussion.} Recurrence \ref{eq:lprd} is valid since $\psi_i$ is a bijection on its domain. Recurrence \ref{eq:leftprop} already delivers a random cyclic permutation at each step. The cut and riffle together shuffle the codomain of that permutation pseudorandomly (like a deck of cards) and the result significantly depends on $p_{i-1}$. The reason why we include the riffle as well as the cut is that we wish to avoid dependency on the sum of $p$'s  as the permutations accumulate with each recurrence step. Without the riffle, at step $i+1$ we would observe the cut at the distance $p_i+p_{i+1}$ as the combined result of two steps for all but two components of the permutation cycle. On the other hand, there will be no consistent similarity between $\pi_i$ and $\pi_{i+1}$ with the riffle. The diffusion will be the greater the more steps have been taken with the recurrence relation, just as the deck of cards will be randomised as the deck is repeatedly cut and riffle-shuffled. This takes care of the requirement 3 above.

\section{Preconditioning}

Recall that the problem we are attempting to solve is that Moriarty's attack strategy can be based on the valid ciphertext that he has intercepted. Even though the redundancy in the plain text makes the acceptance of a random string sampled from the full domain and posing as a ciphertext to a valid plain text exponentially unlikely, the probability distribution in a close vicinity of the valid cipher text can be uneven. Moriarty is in a position to mount a {\em low-dimensional attack} whereby only a few components of the intercepted ciphertext are altered. Those close alterations (in some metric sense) may decipher to a valid plaintext with a higher probability as the example of the OTP presented earlier suggests.  

The template introduced with Proposition \ref{prop:otp} ensures that, as we decrypt, alterations of the ciphertext propagate towards the "big end" in terms of the factoradic number system. This makes a low-dimensional attack on all but the greatest values of $i$ very difficult. To achieve a perturbation confined to some area of the plaintext, Moriarty would have to go through all possible combinations of the ciphertext values in that area. The mapping of the ciphertext on the corresponding values of the plaintext in the affected area is random (if the key is unknown), so full scan through the legal ranges of the Lahmer components is required. However in the process of scanning, all values of $\pi_j$, $j>i$ will be affected as well. This means that the attack cannot be low-dimensional, unless the attack area is flush against the higher end of the interval of $i$, i.e. big-endian.

Indeed, if, say three big-endian components, $\nu-3\le i<\nu$, are given all possible legal values, that is $(\nu-1)(\nu-2)(\nu-3)$ combinations, we are guaranteed that $3!-1=5$ out of them will be deciphered as permutations of the original triplet of plaintext symbols 0,1 and 2. In the process none of the other plaintext components will be affected since alterations propagate only towards the big end and since conversion from Lehmer to one-line representation, although proceeds towards the little end, depends only on the set of the symbols produced previously and not on their relative order. So for example, for $\nu=20$, the codewords
\[
5\ 11\ 7\ <\hbox{tail}>
\]
and
\[
12\ 8\ 5\ <\hbox{tail}>
\]
with the same tail, will produce the same one-line representation for the tail. Indeed in both cases the first three symbols' positions form the {\em set} 
$\{5,12,8\}$ and the tail components will skip over positions from that set. 

This creates the possibility of a low-dimensional attack at a small polynomial cost, an entirely unsatisfactory situation. The attack would have to be on the big-end components of the Lehmer code and so we require an additional diffusion mechanism that propagates towards the little end. 

\subsection{Big-endian Pseudo-Hadamard Transform\label{sec:pht}}

Consider the first $s$ Lehmer components of a size-$\nu$ codeword: $W_0,\ldots,W_{s-1}$. The number of combinations of their legal values is clearly 
\[
Z = \nu\times(\nu-1)\times\ldots\times(\nu-s+1) ={\nu!\over(\nu-s)!} 
\]
now represent $Z$ as the product of its prime factors:
\[
Z = \prod_{j=1}^t f_j^{e_j}
\]
\begin{proposition}\label{prop:crt}
Let $j$ be a natural index and $\nu$ and $s$ be such that \[
(\forall j\le t) f_j^{e_j}<\nu-s+1\,.
\]
Then there exists a bijective map of $W_0,\ldots,W_{s-1}$ on a sequence of nonnegative $r_j$, $j\le t$, where all $r_j<f_j^{e_j}$.
\end{proposition}
\begin{proof}
Observe that for fixed $\nu$ and $s$ the map 
\[
W = \sum_{i=0}^{s-1} {(\nu-i)!\over (\nu-s)!}W_i
\] 
is a bijection. Indeed $W(\nu-s)!$ is the factoradic value of the truncated Lehmer codeword \[
\check{W}_i = \begin{cases}
W_i\,\;\hbox{if}\;i<s\\
0\,\;\hbox{otherwise}
\end{cases}\,
\] 
and as such is convertible to a sequence of factoradic digits (Lehmer code components) and back.
Note that  by construction $0\le W<Z$. 

On the other hand, the Chinese Remainder Theorem (CRT, \cite{CRT}) states that there exists a unique nonnegative $R<Z$ such that 
$r_j = R \mod f_j^{e_j}$ for all $j\le t$. (This is because all $f_j^{e_j}$ are mutually coprime by construction.
Given $R$, we immediately get all $r_j$ and the converse requires a known algorithm with linear complexity.) It
means that there exists a bijection $W\to R$.

We conclude that the combined map $\{W_i\}\to W\to R\to \{r_j\}$ if bijective.
\end{proof}

Proposition \ref{prop:crt} enables a transformation of the original codeword that entangles components $W_0,\ldots,W_{s-1}$ and components $W_{g_1},\ldots,W_{g_t}$, where $g_j = \nu - 1 - f_j^{e_j}$. Consider the following {\em Pseudo-Hadamard} Transform (PHT), introduced in \cite{twofish} as a building block for a symmetric cipher:
\begin{eqnarray}
R^*&=& W+R \mod Z \label{eq:pht1}\\
W^*&=& W+R^* \mod Z\label{eq:pht2}\,.
\end{eqnarray}
The inverse transform is as follows:
\begin{eqnarray}
W&=& W^* - R^* \mod Z\label{eq:pht3}\\
R&=& R^* - W \mod Z\label{eq:pht4}\,.
\end{eqnarray}

This suggests the following enhanced variant of ND-OTP. Assume that $\nu$ and $s$ are fixed; we return to the choice of these 
parameters later. For {\bf encryption}:
\begin{enumerate}
\item Turn the plaintext into a Lehmer codeword using factoradic conversion
\item \label{li:prec}Turn $W_0,\ldots,W_{s-1}$ into $W$ using truncated factoradic conversion (replacing factorial place values by numbers of permutations)
\item Turn $W_{g_1},\ldots,W_{g_t}$ into $R$ using the standard method associated with the CRT \label{item:crte}
\item Obtain $W^*$ and $R^*$ by PHT, Eqs \ref{eq:pht1}-\ref{eq:pht2}
\item Calculate new values $W^*_0,\ldots,W^*_{s-1}$ from $W^*$ using truncated factoradic conversion and update the Lehmer codeword accordingly
\item \label{li:precf} Obtain new values $W^*_{g_1},\ldots,W^*_{g_t}$ from $R^*$ by calculating the Chinese remainders, and update the Lehmer codeword accordingly
\item Apply ND-OTP encryption to the updated plaintext codeword and obtain the ciphertext codeword, then convert from factoradic to binary
\end{enumerate}

We call steps \ref{li:prec}--\ref{li:precf} {\em preconditioning} the plaintext. The purpose of preconditioning is to make big-end alterations impossible without altering Lehmer components from the lower-half of the codeword as well. A preconditioned plaintext contains the same information as the original one; it is only the representation that is different.

For {\bf decryption} the method is as follows:
\begin{enumerate}
\item Convert the ciphertext from binary to factoradic. Apply ND-OTP decryption to the ciphertext codeword. Obtain the preconditioned plaintext
\item Turn $W^*_0,\ldots,W^*_{s-1}$ into $W^*$ using truncated factoradic conversion
\item Turn $W^*_{g_1},\ldots,W^*_{g_t}$ into $R^*$ using the CRT method \label{item:crtd}
\item Obtain $W$ and $R$ by PHT, Eqs \ref{eq:pht3}-\ref{eq:pht4}
\item Calculate original values $W_0,\ldots,W_{s-1}$ from $W$ using truncated factoradic conversion and update the Lehmer codeword accordingly
\item Obtain original values $W_{g_1},\ldots,W_{g_t}$ from $R$ by calculating the Chinese remainders, and update the Lehmer codeword accordingly
\item Turn the resulting Lehmer codeword into the original plaintext by converting factoradic to binary
\end{enumerate}

\subsection{Configuration}
The choice of $\nu$ and $s$ has been left out. We will discuss how this choice affects the nondegeneracy of the cipher in the next subsection. But first let us focus on quantifying the parameters.

Recall that Proposition \ref{prop:crt} has the following condition. For any choice of $\nu$ and $s$ there exist a positive $t$ and prime numbers $f_j$, $j=1,\ldots,t$ such that 
\begin{equation}
\nu\times(\nu-1)\times\ldots\times(\nu-s+1) = \prod_{j=1}^t f_j^{e_j}\,,\label{eq:decomp}
\end{equation}
where all $e_j$ are natural numbers. This factorisation is unique up to the ordering of $f_j$. The choice of $\nu$ and $s$ is valid only when 
\begin{equation}
(\forall j\le t) f_j^{e_j}<\nu-s+1\,.\label{eq:crtx}
\end{equation}
Obviously for every valid pair $(\nu,s)$ the pair $(\nu,s^\prime)$, where $1<s^\prime<s$ is also valid, so we are interested in the greatest $s$ for each given $\nu$. Also observe that if the interval $[\nu,\nu-s+1]$ contains a power of a prime, then it must be one of the factors $f_j^{e_j}$, and so the condition in Eq \ref{eq:crtx} is broken. With this in mind, we have performed a direct search for suitable parameters for $\nu<200$ (message sizes up to, or around, 2 Kb), see Table \ref{tab:fact}, noting the first values of $\nu$ for which $s_{\rm max}$ leaps up.

\begin{table}
\begin{center}
\begin{tabular}{|c|c|l|l|}
\hline
$n$&$\nu$ & $s_{\rm max}$ & $f_j^{e_j}\vphantom{3^{3^{3^3}}}$\\
\hline
69 & 22 & 3 & 3, 5, 7, 8, 11 \\
138 & 36 & 4 & 5, 7, 8, 11, 17, 27 \\
382 & 78 & 5 & 7, 9, 11, 13, 16, 19, 25, 37 \\
491 & 95 & 6 & 7, 13, 16, 19, 23, 25, 27, 31, 47 \\
851 & 147 & 7 & 5, 11, 13, 29, 47, 49, 64, 71, 73, 81 \\
1299 & 207 & 8 & 7, 17, 23, 29, 41, 67, 81, 101, 103, 125, 128 \\
2066 & 303 & 9 & 7, 11, 13, 23, 37, 43, 59, 101, 125, 128, 149, 151, 243 \\
\hline
\end{tabular}
\caption{Factorisation data for the least $\nu$ of a given $s_{\rm max}$. The first column shows the corresponding plaintext/ciphertext size in bits \label{tab:fact}}
\end{center}
\end{table}

\subsection{Discussion}

A preconditioned plaintext makes the big-endian attack infeasible even for a small $s$. The attacker would have to try all possible combination of the first $s$ components of the Lehmer codeword at the big end, since the mapping of the components of the ciphertext on the corresponding components of the preconditioned plaintext is, although bijective, still random thanks to the encipherment according to Proposition \ref{prop:otp}. Since it is a bijection, if all combinations are attempted in the ciphertext, then after decryption all combinations of the preconditioned plaintext will be seen, too, if only in a different order. As the preconditioning is reversed, all the original values of the first $s$ Lehmer components will be covered, as follows from Eq \ref{eq:pht3}. However, as evidenced by Eq \ref{eq:pht4}, at the same time all values of the Chinese remainders will also be seen in some order. Since the positions $\{f_j^{e_j}\}$ are scattered over the little-end half of the Lehmer codeword, and they are not dense in it, the reconstruction of the original sequence of symbols will potentially affect all symbols from the left-most position of the set down. For example, as evidenced by Table \ref{tab:fact}, for $\nu=95$, scanning through the first six components of the ciphertext would potentially affect all components from 47 to 1, which is half of the Lehmer word size. To stop this happening, Moriarty could also scan through the values in the position set $\{f_j^{e_j}\}$ to try and find a combination that results in a close (in the metric sense) permutation of the original. This, however, would not work, since the cipher is left-propagating: a change of any values in the position set will affect the decryption of all the subsequent values towards the big end, thus destroying the localised character of the attack.

For ease of reference we provide the formula for step \ref{item:crte} of the encryption and decryption sequences, based on the Extended Euclid Method\cite{CRT}. For encryption, the computation of R is as follows:
\begin{equation}
R=\sum_{i=1}^{t}W_{g_i}b_ib^\prime_i \mod Z\label{eq:euclid}
\end{equation}
where
\[
b_i = Z/f^{e_i}_i
\]
and
\[
b^\prime_i = b^{-1}_i \mod f^{e_i}_i\,,
\]
and similarly for $R^*$ for decryption. Note that $b^\prime_i$ exists and is unique due to $b_i$ and $f^{e_i}_i$ being coprime by definition. Also note that all $b_ib^\prime_i$ in Eq \ref{eq:euclid} can be precomputed for the choice of $\nu$ and so Step 3 involves $t$ modulo multiplications and $t-1$ additions and has a sublinear complexity in $\nu$.

Yet one question remains. The minimum alteration of a component that belongs to the position set is 1. This follows from the fact that by scanning through the $s$ big-end components all combinations of the Chinese remainders are enumerated. In particular, there will be a combination of big-endian values that results in the same remainders as those for the original plaintext, except the value in the position closest to the little-end. In the case $\nu=95$ this would be position 7. In particular the alteration of the value in that position from the original plaintext could be by any value less than 7, including 1, which results in at most a single transposition of symbols\footnote{Recall that the plaintext is a Lahmer codeword, not the eventual symbols that it encodes}. We conclude that the proposed scheme still leaves open an opportunity of an attack with no more than $s+1$ transpositions. Even though the statistical weight of such perturbations is small, we should seek a stronger diffusion mechanism that works toward the little end, which is our next step.

\subsection{Differentiating the Lehmer code\label{sec:fder}} 

\begin{figure}[t]
\begin{center}
\includegraphics[scale=1.00]{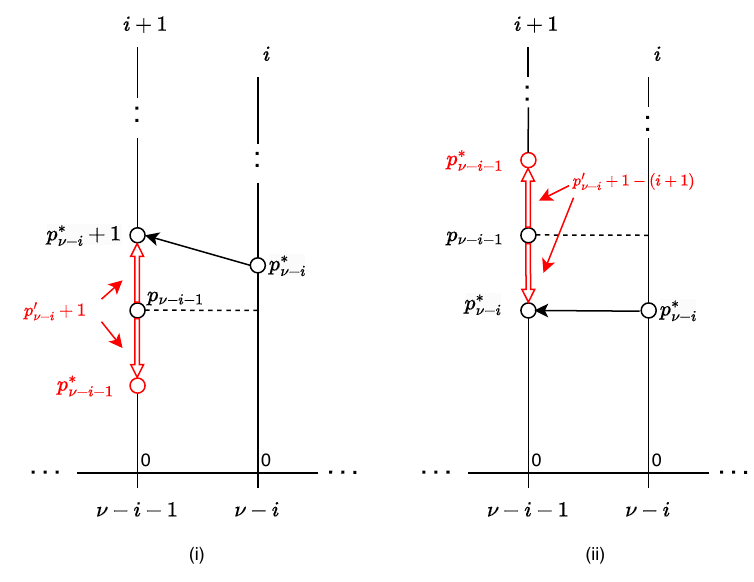}
\caption{One step of the recurrence relation in Eqs \ref{eq:dif1}-\ref{eq:dif2}}
\label{fig:dproof}
\end{center}
\end{figure}

\begin{definition} Let $p$ be a permutation of  $0,\ldots,\nu-1$ and let $\langle p_i\rangle$ be its Lehmer codeword. The first derivative 
$\langle{p}^\prime_i\rangle$ of the permutation is defined by the following recurrence relation\/{\rm :}
\begin{equation}
p^\prime_{\nu-1}=p_{\nu-1}=0,\;\; p^*_{\nu-2}=p_{\nu-2}\;;\label{eq:init1}
\end{equation}
for $i=2,\ldots,\nu-1${\rm :}
\begin{eqnarray}
p^\prime_{\nu-i} &=& p^*_{\nu-i}-p_{\nu-i-1} \mod i \label{eq:dif1}\\
p^*_{\nu-i-1}&=& 
p_{\nu-i-1} - p^\prime_{\nu-i}-1 \mod i+1\label{eq:dif2}
\end{eqnarray}
\noindent Finally,
\begin{equation}
p^\prime_0=p^*_0\label{eq:init2}
\end{equation}
\end{definition}
\noindent Note that the first derivative satisfies the Lehmer code constraint:\[
0\le p^\prime_i<\nu-i\;\;\; \text{for all } 0\le i<\nu-1,
\]
and so it is a Lehmer codeword in its own right. This makes it possible to combine pre-differentiation with NDOTP.
\begin{proposition}
The mapping of a Lehmer codeword $\langle p_i\rangle$ onto its first derivative $\langle p^\prime_i\rangle=\mathcal{D}\left(\langle p_i\rangle \right)$ is bijective. The inverse mapping $\mathcal{D}^{-1}$ is given by the following recurrence relation\/{\rm:}
\begin{equation}
p^*_0=p^\prime_0\,;\label{eq:init3}
\end{equation}
for $i=0,\ldots,\nu-3${\rm:}
\begin{eqnarray}
p_{i} &=& p^*_i+p^\prime_{i+1}+1 \mod \nu-i\label{eq:dif3}\\
p^*_{i+1} &=& p_i+p^\prime_{i+1} \mod \nu-i-1\label{eq:dif4}
\end{eqnarray}
\noindent Finally,
\begin{equation}
p_{\nu-2}=p^*_{\nu-2}\label{eq:init4}
\end{equation}
\begin{equation}
p_{\nu-1}=0\label{eq:init5}
\end{equation}
\end{proposition}
\begin{proof}
Substitute $j$ for $i$ in Eq \ref{eq:dif1} and $j$ for $\nu-i-1$ in Eq \ref{eq:dif4}; they will become identical. Use the same substitutions with Eqs \ref{eq:dif2} and \ref{eq:dif3}, respectively, and they, too, will become identical. The  boundary conditions Eqs \ref{eq:init1}, \ref{eq:init2} are consistent with those given by Eqs \ref{eq:init3}, \ref{eq:init4} and \ref{eq:init5}.
\end{proof}

\paragraph{Discussion.} The diagram in Figure \ref{fig:dproof} illustrates the relationship between the current and next state of the recurrence relation defined by Eqs \ref{eq:dif1}-\ref{eq:dif2} and helps to understand the properties of the proposed derivative. It is convenient to consider two separate cases, $p^*_{\nu-i} \ge p_{\nu-i-1}$ (case (i)), and $p^*_{\nu-i} < p_{\nu-i-1}$ (case (ii)), even though they are captured by the same pair of equations.  

We start with the first case, and treat the Lehmer component for the symbol $\nu-i$ as a position of the symbol in the sequence taking into account the potential skip over the symbol $\nu-i-1$, which happens if $p^*_{\nu-i} \ge p_{\nu-i-1}$, as is the case here. To calculate the position difference, symbol $\nu-i$ should be placed in the domain of the symbol $\nu-i-1$, which requires incrementing the Lehmer component of the former. Next we determine the distance by travelling in the positive direction from the lesser symbol towards the greater one, as the red arrow indicates. Clearly, since the image of $p^*_{\nu-i}$ in the domain of $\nu-i-1$ never collides with $\nu-i-1$ and since there are $i+1$ positions in total (from 0 to $i$) available to the symbols, the distance between them is in the range $[1,i]$. If we subtract 1 from the distance, we can assign the result to the derivative component $p_{\nu-i}$, which takes values from the range $[0,i-1]$. The procedure is consistent with Eq \ref{eq:dif1}. Finally we reverse the direction of the distance vector and shift $p_{v-i-1}$ down, Eq \ref{eq:dif2}. The value $p^{\prime}_{\nu-i}$ is the outcome of differentiation, and $p^*_{\nu-i-1}$ represents the next state of the recurrence.

Case (ii) is similar, except $p^*_{\nu-i} < p_{\nu-i-1}$ and so there is no skip. We do not increment $p^*_{\nu-1}$ when we map it on the domain of the symbol $\nu-i-1$. The distance is now negative and we must add $i+1$, the modulus for the symbol $\nu-i-1$, and only then, for reasons exposed with case (i) we also subtract one from the result. We arrive at the same equation, Eq \ref{eq:dif1}. The shift that produces the next state now happens in the opposite direction as well, which is in accordance with Eq \ref{eq:dif2}.

The above diagrams also help to understand the integration defined by Eqs \ref{eq:dif3}-\ref{eq:dif4}, but more importantly, it shows the sensitivity of differentiation. Substituting Eq \ref{eq:dif3} in Eq \ref{eq:dif4} we obtain 
\begin{equation}
p^*_{i+1} =  (p^*_i+p^\prime_{i+1}+1 \mod \nu-i)  +p^\prime_{i+1} \mod \nu-i-1\label{eq:diftot}
\end{equation}
from which we can see that $p^*_{i+1}$ depends on both $p^*_i$ and $p^\prime_{i+1}$. A change in $p^\prime_j$ will result in a change in $p^*_j$ for any given $j$: the red arrows on the diagram are of a length less than the modulus $i+1$ by construction. Furthermore, a change in $p^*_j$ will result in a change in $p^*_{j+1}$ even if $p^\prime_{j+1}$ remains the same. Changes in $p^*$ will propagate towards the little end (high indices) affecting all $p$'s along the way. This may seem like a much stronger diffusion mechanism than the preconditioning, described in Section \ref{sec:pht}. However, many  combinations of $p_i$ with $i<j$ for some $j$ sufficiently greater than 0 will result in the same $p^*_j$, and then it follows from Eq \ref{eq:diftot} that all $p_{i>j}$ will remain the same for each of those combinations, thus enabling a low-dimensional attack. Only the combined effect of the differential plaintext and the Pseudo-Hadamard Transform on the big-endian components can make the big-endian attack impractical. 

For example, for the case $\nu=95$, $s=5$, see Table \ref{tab:fact}, the attacker will have to try every combination of the first seven ciphertext components. Due to the left-propagation of the decryption process, no modification of the remainder positions 7, 13, 16, $\ldots$ or their vicinity is possible since that would affect the decryption of all positions towards the big end, and the integration process will propagate to the little end the new values of the remainders changed as a result of the Pseudo-Hadamard Transform of the changed big-end values. This means that the last 47 positions will all be affected in addition to the directly affected seven (the six in the transform plus another one to stop the run-away integration process) at the big end. 

\begin{figure}[t]
\begin{center}
\includegraphics[scale=0.6]{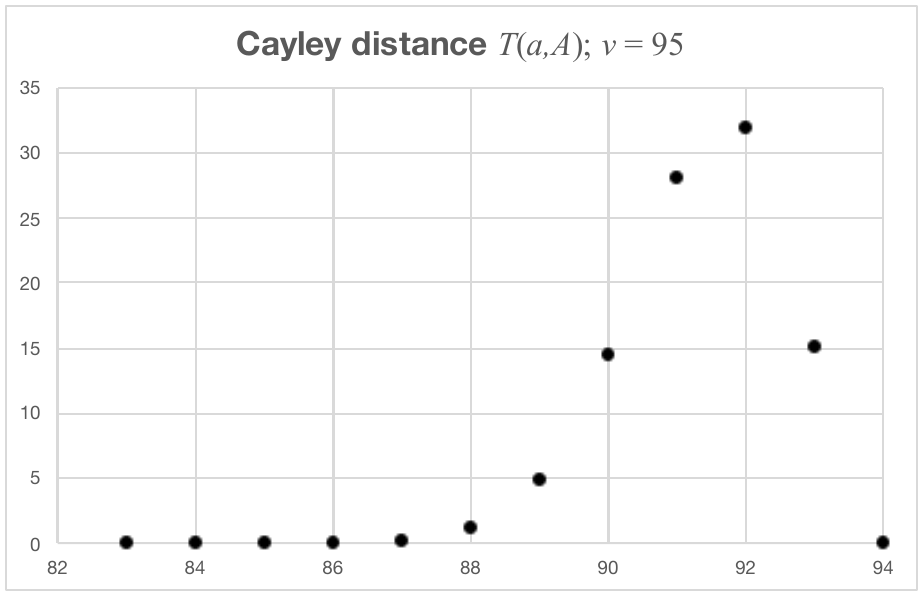}
\caption{Histogram of the Cayley distance from the perturbed to original $a$ obtained by Monte Carlo method. The computed mean $\bar{T}(a,A)\approx90.84$, in keeping with the expected $\nu-\ln\nu\approx90.45$}
\label{fig:diff-metric}
\end{center}
\end{figure}

To assess the diffusion properties of differentiation, recall from the metric theory of the symmetric group that the average Cayley distance between two random permutations of a length-$\nu$ sequence is $\nu-\ln\nu$ \cite{metrics-diaconis}. Let us take the first derivative of an arbitrary permutation $a^\prime$, alter the first element of its sequence at random, and integrate the result to obtain some permutation $A$. We get an ensemble of $A$, in which we determine the Cayley distance (the minimum number of transpositions required to reach the destination) from each to $a$, $T(a,A)$. Now calculate the mean distance and average it over $a$. If the diffusion from differentiating is good we should get the average distance in that ensemble close to the average distance in the whole group. Figure \ref{fig:diff-metric} shows the histogram of $T(a,A)$ we obtained by Monte Carlo simulation for $\nu=95$. Although a crude measure, the mean distance is very close to that between a pair of permutations chosen at random, which indicates a good measure of diffusion. The choice of metric (the Cayley distance) is not unique, since several others are available (the more familiar Hamming distance, widely used in cipher analysis, among them). However, our proposed method of redundancy injection (see next Section) is sensitive to transpositions and so a transpositional distance seems the most pertinent. 

\section{Injecting Redundancy}

In the previous sections we developed an NDOTP with random diffusion towards the big end and a preconditioning that helps to diffuse towards the little end a big-endian attack on the ciphertext, especially if the plaintext is pre-differentiated. Those measures should pave the way to application of standard redundancy mechanisms within the plain text that are robust enough with respect to an unlocalised random perturbation. Such redundancy could be introduced in the original binary message using a simple linear code. However, once the factoradic conversion has been made, redundancy can also be injected in the target representation. As it turns out, a function exists that requires nothing more than a lookup table for its implementation, and whose mapping is robust enough to withstand even the most localised attack, if only with a slightly increased redundancy overhead. We will define this function next.

\subsection{Pseudo Foata Injection}

Dominuque Foata is known for his extensive contribution to combinatorics, among which there is the so-called Foata bijection\cite{Foata}. That bijection is a map of the symmetric group onto itself (not a morphism) based on a ``pun'', i.e. deliberate misreading of the cyclic representation of a permutation as the one-line one. Foata takes credit for finding that under some ordering of the cycles, the pun is invertible and the map is a bijection. We use Foata's pun to create an {\em injection} to the next order of the group, rather than a bijection.

\begin{definition}\label{def:pfi}
For a natural $n$, the Pseudo Foata Injection (PFI) is a map ${\FF}_n:S_n\to S_{n+1}$ defined as follows.
Let $\langle p_i \rangle_{0\le i <n}$ be the one-line representation of a permutation $p$ from $S_n$. Then
${\FF}_n(p)$ is the permutation $q$ whose {\bf cyclic} representation consists of a single cycle 
\[(p_0,\ldots,p_{n-1},n)\,.\]
\end{definition}
\noindent The injective property of the PFI is obvious from its definition. The one-line representation of $q$, $\langle q_i \rangle_{0\le i <n+1}$ follows directly from Definition \ref{def:pfi} as well:
\[
q_i = 
\begin{cases} 
p_{m+1} &\text{if } (\exists m<n-1)p_m = i\\ 
n, &\text{if } p_{n-1}=i,\\ 
p_0, & \text{otherwise.}
\end{cases}
\]

\begin{definition}\label{def:ipfi}
A partial function ${\FF}^{-1}_{n+1}:S_{n+1}\to S_{n}$ is defined on the subset of $S_{n+1}$ that comprises all single-cycle permutations $q$. Since the cyclic representation of $q$, $(q_0,\ldots,q_n)$ is ambiguous and we can assume without loss of generality that $q_n=n$ {\rm(}\/if not, rotate the cycle until the equation holds\/{\rm)}. Function ${\FF}^{-1}_{n+1}$ maps every $q$ on the $p$ whose {\bf one-line} representation is $q_0,\ldots,q_{n-1}$. 
\end{definition}
\begin{proposition}
For all $p\in S_n$, ${\FF}^{-1}_{n+1}({\FF}_{n}(p))=p$.
\end{proposition}
\begin{proof}
Follows from Definitions \ref{def:pfi},\ref{def:ipfi}.
\end{proof}
\begin{corollary} 
For any $n,k>0$, $p\in S_n$, ${\FF}^{-k}_{n+k}({\FF}^k_{n}(p))=p$
\end{corollary}

\begin{proposition}
The density of the image of ${\FF}_n$ in its codomain is $1/(n+1)$.
\end{proposition}
\begin{proof}
The PFI is an injection, so the density is the domain to codomain ratio: $|S_n|/|S_{n+1}|=1/(n+1)$ 
\end{proof}
\noindent The following are corollaries of the above for repeated application of the PFI:
\begin{corollary}\label{cor:dens}
The density of the image of ${\FF}^k_n$ in its codomain is $n!/(n+k)!$
\end{corollary}
\noindent In the above corollaries 
\[\FF^k_n(p) = \FF_{n+k-1}(\ldots \FF_{n+1}(\FF_{n}(p)\ldots))\] 
and 
\[\FF^{-k}_{n+k}(p) = \FF^{-1}_{n+1}(\ldots \FF^{-1}_{n+k-1}(\FF^{-1}_{n+k}(p)\ldots))\]
denote repeated applications of the function $\FF$ and its inverse.

Let us set $\nu=85$ as an illustration and consider the redundancy injection by $\FF^{10}_\nu$ to obtain a plaintext for the NDOTP sized 95, which we use as a running example. Using to Corollary \ref{cor:dens} and assuming the attack results in a random deviation from the plaintext after deciphering (de-preconditioning, etc), we conclude that the success probability of the attack is $85!/95!<2.8\times 10^{-20}$ which is the probability expected from a 64-bit redundancy code. 

As a concluding remark, we wish to mention that the complexity of the PFI is linear in $n$ and the implementation only uses table lookups and table storage, so it is probably the fastest way of injecting redundancy in a message if the message is already in the form of a permutation in the one-line representation. With NDODT, the binary message is turned into factoradic/Lehmer first; to utilise PFI the Lehmer code would have to be converted to one-line representation (at a quadratic cost) and then back (also at a quadratic cost) for the message with redundancy to be enciphered. Similar conversions would have to be made after deciphering.

\subsection{Statistics of the PFI}

We have discovered that despite its simplicity, PFI is quite robust in resisting the low-dimensional attacks, i.e. the type of attack that we introduced pre-conditioning and differentiation to thwart. To obtain a statistical illustration we performed a Monte–Carlo experiment in which we  
\begin{enumerate}
\item used 100,000 random 50-symbol plaintexts and injected them with 10 more symbols using $\FF^{10}_{50}$
\item each of the 100,000 60-symbol results was subjected to all possible perturbations with a Cayley distance of 2 using the 
transposition pattern $abc\to bca$ exhaustively
\item attempted an appropriate $\FF^{-1}$ on each transposition result repeatedly until the result became acyclic. The {\bf penetration depth}, i.e. the number of successful applications of $\FF^{-1}$ before the result becomes acyclic was histogrammed.
\end{enumerate}

\begin{figure}
\begin{center}
\includegraphics[scale=0.8]{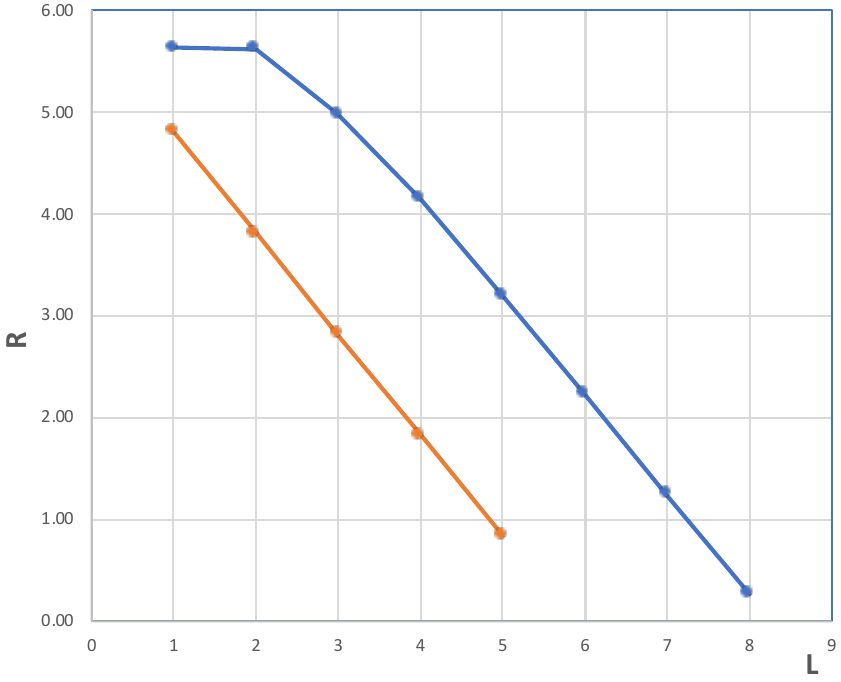}
\caption{Blue: penetration rate $R$ vs penetration depth $L$. The vertical axis is in logarithmic scale (base 60). Red: the same for random perturbation (Corollary \ref{cor:dens})}
\label{fig:pendepth}
\end{center}
\end{figure}

In point 2 above the reason for using distance 2 is that a single transposition (Cayley distance 1) breaks a single cycle into two which makes $\FF^{-1}$ undefined on the result. The chosen pattern is the smaller perturbation than the other one available at distance 2, i.e. $(ab\to ba, cd\to dc)$ in terms of another important metric, Hamming, and so represents the smallest possible perturbation of the sequence. More than  $2\times10^{10}$ outcomes in total were collected. The results are displayed in figure \ref{fig:pendepth}.

The penetration statistics are placed on the logarithmic scale using base 60, so that the vertical drop of 1 unit represents the reduction by a factor of 60. If completely random permutations were presented to the inverse PFI, one would expect the curve to decline nearly linearly, in accordance with Corollary \ref{cor:dens}; see the red curve. Since we deliberately tried the closest possible permutation to the one that actually satisfies the tenth power of $\FF^{-1}$, we observed a much higher penetration rate, a factor of $60^2$ higher. As follows from the plot, increasing the number of PFI applications by 3, i.e. going from 5- to 8-symbol redundancy levels off the difference in the worst case scenario of the low-dimensional attack. To put things in perspective, after $2\times10^{10}$ attempts on behalf of Moriarty we have not seen a single outcome at depth 9; we must have been two orders of magnitude away in the number of attempts from successful penetration. 

\section{Related work\label{sec:relwork}}

One of the earlier works where secrecy and integrity were approached as separate concerns even for a perfect cipher was by Desmedt \cite{desmedt}. However the seminal work in the area of unconditional integrity was published by Carter and Wegman \cite{U2}, where the method was proposed based on a family of so-called {\bf universal}$_2$ hash functions. A message to be sent consists of the message proper $m\in M$ and its integrity tag $t\in T$. The family $H$ is of functions $M\to T$ and it has the property that for any messages $m_1\ne m_2$, 
\[
{|\{ h\in H \mid h(m_1) = h(m_2)\}|\over |H|} \le 1/|T|\,.
\]
\noindent The family $H$ is typically indexed by a key $K$, $H=\langle h_K \rangle_{K\in \{0,1\}^L}$. 

Later on \cite{strongU2}, a more restrictive class of families was defined, called strongly universal$_n$. Such a family contains exactly $|H|/|T|^n$ functions that map any $n$ pairwise distinct messages $a_1,\ldots,a_n$ onto any $n$ (not necessarily pairwise distinct) tag values $t_1,\ldots,t_n$. Also, the fixed security parameter $1/|T|$ has been generalised to an arbitrary $\epsilon$ giving rise to $\epsilon$-Almost Strongly Universal$_2$ hash functions. But let us return to the integrity assurance.

The idea is that Alice and Bob share a secret $K$ chosen at random. Alice then uses $h_K$ from a pre-agreed strongly universal family to produce an authentication tag $h_K(m)$, enciphers the tag using an OTP and sends it to Bob along with the message $m$. Bob produces his own tag value using the received message $m^\prime$, $t^\prime= h_K(m^\prime)$ and compares it with what Alice sent, after deciphering. Although Moriarty can intercept and alter $m$ to $m^*$, he can only guess $K$, since $t$ is perfectly secret (whereas $m$ may not be secret at all). Moriarty is thus unable to compute the new tag $t^*$, but if he chooses it at random, strong universality$_2$ limits his success probability to $1/|T|$.    

This avenue of research is still active even though the early works were published more than 40 years ago. Various families of universal hash functions continue to be constructed and their applications to cybersecurity stretch as far as QKD protocols (e.g. see a recent paper \cite{QKD-hash}). Since universal hash families provide purely statistical guarantees (they are {\em unconditional}) they will work with standard binary OTPs. 

Implementation of universal families generally follows two methods. One, which was proposed in the very first publications on the matter, is number-theoretical. The hash value is achieved as an affine product of the key and the message, both represented as vectors with components modulo prime number. There are number-theoretical reasons why the result is well spread over the tag space, which makes the construction strongly universal. The other method was proposed in \cite{LFSR}, and it is based on a random matrix, produced from a key with the help of a linear feedback shift register. Both methods originally required a key that was longer than the message itself. However, new structures were put forward, which reduced the key length to just over four times the tag length $n$, with the latter defining the probability of forgery as $2^{-n}$. For more details, see \cite{newU2}, where Table 3 summarises the known methods in terms of their security parameter and key length.

There is an important difference between integrity provided by universal hashing and our approach. The statistical guarantee of the former comes from a physically random key that injects entropy in the tag. Our proposal, the Pseudo Foata Injection, does not rely on the entropy of the mapping parameter for its effect. Instead, by making the cipher non-degenerate, Moriarty is denied the chance to make localised (we call them low-dimensional) alterations to the redundant plaintext. As a result, any change in the intercepted ciphertext results in random (due to the OTP's random key), unpredictable (due to the Pad's non-degeneracy) changes to both the "message" and its "tag"; those are inseparable and well spread-out in the PFI output. As a result, the probability of Moriarty's success is defined solely by the cardinality ratio of the PFI range to codomain, just like the security of the universal hash family is determined by a similar cardinality ratio: the set of colliding hash functions to the full set. 

The material gain of our approach is the lack of a random "integrity key", which, as we mentioned could reach four times the amount of redundancy injected by the tag. I our case, the injected redundancy is according to Corollary \ref{cor:dens}, but the only extra key material that we need for it is that for the lengthened OTP key to cover the longer message. OTP protection of the tag would require the same in addition to the integrity key.   

\section{Conclusions}

We have presented four tools from a toolkit that provides perfectly secret messages with an integrity assurance. The primary result of this work is the construction of a non-degenerate OTP, Section \ref{sec:ndotp}, which obviates the need for randomisation of message redundancy. The key advantage of the proposed NDOTP is the fact that it can use the encryption key without constraining it (and thus destroying the perfect secrecy) for a second purpose: to entangle the encipherment of units of the plaintext. Normally the key cannot be used twice, since the trijection requirement would necessitate the consistency of both uses for any pair of plain- and ciphertexts. However, the ability of any cyclic subgroup of permutations to deliver a given symbol in any given place of the sequence decouples the two uses. The entanglement, however, only works in one direction (little- to big-end), and so the diffusion introduced by the non-degeneracy is not all-with-all. However, the nature of the entanglement is random as it is based on the encryption key. The latter has to be physically random, i.e. have maximum entropy, to prevent information leakage from plain- to ciphertext, and we are able to utilise that entropy in the entanglement. Another key advantage  is nice differential behaviour of NDOTP: even at the big end, where there is no diffusion to the left, it is impossible to predict the effect of a small change of the ciphertext on the plaintext: the generator of the cyclic subgroup that does that mapping is dependent on the random key. As a result, a small deviation of the plaintext can only happen in the course of exhaustive search through the ciphertext components of interest.  

To address the unidirectional diffusion property of the proposed encryption, we introduced two further mappings. One is based on PHT/CRT, see Section \ref{sec:pht}, and it entangles a number of big-end components of the plaintext with some of its components in the little-end part. The latter ones cannot be profitably altered by Moriarty by altering the corresponding components of the ciphertext: all plaintext components to the left of the affected area will be potentially altered as well in the process of deciphering the message. Yet there are only a small number of little-end components affected by any alteration of the big-end ones, see Table \ref{tab:fact}. To make the diffusion to the right stronger, we proposed a further mapping: that of the plaintext on its first derivative, Section \ref{sec:fder}. Differentiating the Lehmer code has a useful side-effect of entangling the little-end components involved in the PHT/CRT as well as any components in between and to the right of them. In our Monte-Carlo experiments we found that differentiation makes the plaintext strongly dependent on its derivative: an alteration of the leftmost component of the derivative results in the change of the plaintext by a distance close to the average distance between members of the symmetric group. In other words, differentiation makes the plaintext extremely sensitive to alterations in even one of its components. The sensitivity is directional: the affected components are on the right of the affected area.   

Finally, we proposed a method of injecting redundancy. Whereas all the above processing is done in the Lehmer representation, the Pseudo Foata Injection requires the one-line representation, which can be obtained from the Lehmer one at a quadratic cost. This should not be a problem, since the factoradic conversion and the NDOTP encipherment/decipherment are also of quadratic complexity. Note that the conversion from the Lehmer to the one-line representation introduces useful diffusion to the right, see Section \ref{sec:Lehmer} and Figure \ref{fig:lemdif}.

PFI is a linear-complexity procedure and it does not involve computation beyond table lookups. We have evaluated its statistical properties on the ensemble of perturbations that are extremely unlikely to be at Moriarty disposal in the light of the properties of PHT/CRT and differentiation discussed above: minimum deviation from the valid redundancy-injected message. Yet we discovered that even in those circumstances PFI is quite robust: it is capable of rejecting ``near'' alterations at the expense of just a few additional redundant components (3 in our case). 

Looking back at the above results, we believe we can make the following conclusions.

The four mappings: NDOTP, PHT/CRT, Differentiation, and PFI constitute a toolkit for integrity assurance of perfectly secret messages. Used separately and in combination they make it possible to find a solution with the right latency and resource footprint for individual circumstances: 

\begin{enumerate}
\item NDOTP can be combined with PFI directly, but the redundancy parameter $k$ should be increased to resist the big-endian attack. Our experiments show that a small increase is sufficient, but further data are needed to tighten the requirements.
\item PFI can be used with other forms of encryption to provide the benefit of distributed redundancy so that Moriarty is unable to attack the "tag" and the "message" separately
\item When preconditioning, NDOTP and PFI are combined, we enjoy the most economic form of perfectly secret communication: the length of the key is equal to the sum of the message length and injected redundancy with all algorithms having no more than quadratic complexity; the result has perfect secrecy and unconditional integrity with a predictable probability of forgery.     
\end{enumerate}
 
\paragraph{Acknowledgement} The author is indebted to Bruce Christianson, who has read the manuscript and made many interesting comments.

\bibliographystyle{plain}
\bibliography{noteref}

\end{document}